\documentclass{aa}

\usepackage{graphicx}

\begin{document}

\title{\bf RR Lyrae variables in Galactic globular clusters}
\subtitle{\bf I: The observational scenario}

\author{M. Castellani \inst{1}  \and  F. Caputo \inst{1}  \and  V. Castellani  \inst{1}$^,$\inst{2}}
\offprints {M. Castellani, \email {mkast@mporzio.astro.it}}
\institute{INAF - Osservatorio Astronomico di Roma, via Frascati
33, 00040 Monte Porzio Catone, Italy \and INFN Sezione di Ferrara,
via Paradiso 12, 44100 Ferrara, Italy}

\date{Received ; accepted }




%



\abstract{In this paper we revisit observational data concerning
RR Lyrae stars in Galactic globular clusters, presenting frequency
histograms of fundamentalized periods for the 32 clusters having
more than 12 pulsators with well recognized period and pulsation
mode. One finds that the range of fundamentalized periods covered
by the variables in a given cluster remains fairly constant in
varying the cluster metallicity all over the metallicity range
spanned by the cluster sample,  with the only two exceptions given
by M15 and NGC6441. We conclude that the width in temperature of
the RR Lyrae instability strip appears largely independent of the
cluster metallicity. At the same time, it appears that the fundamentalized
periods are not affected by the predicted variation of  pulsators
luminosity with metal abundance, indicating the occurrence of a
correlated variation in the pulsator mass. We discuss mean periods
in a selected sample of statistically significant "RR rich"
clusters with no less than 10 RRab and 5 RRc variables. One finds
a clear evidence for the well known Oosterhoff dichotomy in the
mean period $<Pab>$ of ab-type variables, together with a
similarly clear evidence for a constancy of the mean
fundamentalized period $<Pf>$ in passing from Oosterhoff type II
to type I clusters. On this basis, the origin of the Oosterhoff
dichotomy is discussed, presenting  evidence against a
strong dependence of the RR Lyrae luminosity on the metal content.
On the contrary, i) the continuity of the mean fundamentalized
period, ii) the period frequency histograms in the two prototypes
M3 (type I) and M15 (type II), iii) the relative abundance of
first overtone pulsators, and iv) the observed difference between
mean fundamental $<Pab>$ and fundamentalized periods $<Pf>$, all
agree in suggesting the dominant occurrence of a variation in the
pulsation mode in a middle region of the instability strip (the
"OR" zone), where variables of Oosterhoff type I and type II
clusters are pulsating in the fundamental or first overtone mode,
respectively.

}

   \maketitle


\section{Introduction}

Since the pioneering paper by Oosterhoff (1939),  pulsation
periods of RR Lyrae variables in Galactic globular clusters have attracted
the attention of several astronomers because of some clear
evolutionary signatures. Within the restricted sample of clusters
with a rich RR Lyrae population one finds a clear difference
in the period distribution of the pulsators. In particular, the
mean period of the fundamental (RRab) pulsators follows the so
called Oosterhoff dichotomy, grouping below (Oosterhoff type I: OoI)
or above (Oosterhoff type II: OoII)
the value $<Pab>$=0.6 day, or $<$log$Pab>=-$0.22. There is now a
general agreement on the evidence that the dichotomy appears
correlated with the cluster metallicity, the more metal poor
clusters (OoII) having larger mean periods than moderately metal
rich (OoI) ones. However, the origin of such a behavior has raised
a long standing debate which remains unsettled.

\begin{figure*}
\centering
\includegraphics[width=16.0cm]{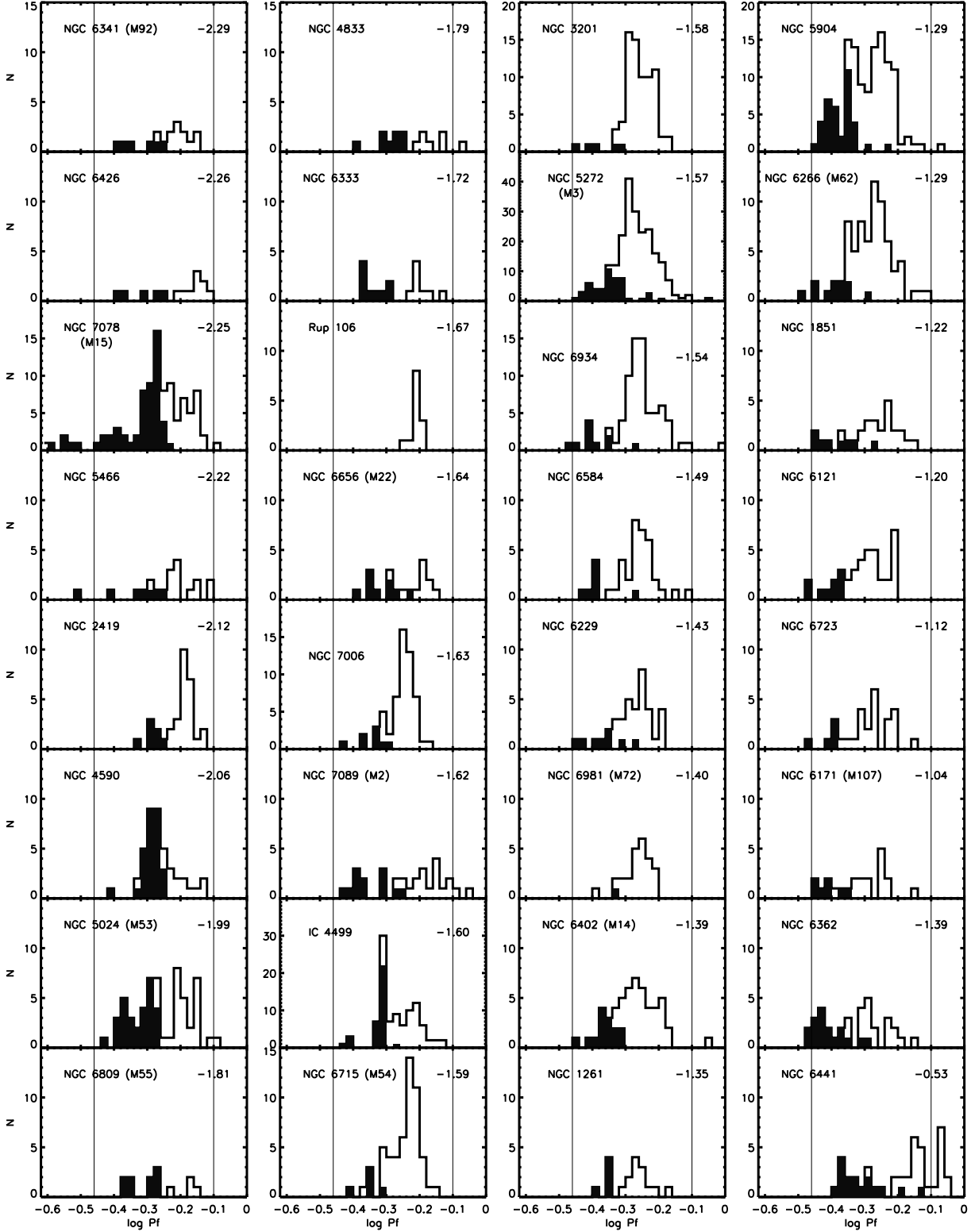}
\vspace*{0.4truecm} \caption{Period frequency histogram for the
selected cluster sample. White and black bars give the
contribution of fundamental and fundamentalized first overtone
pulsators, respectively.}
\end{figure*}

The first pioneering attempt to link stellar evolution to
pulsation was given by Rood (1973), whose synthetic Horizontal
Branch (SHB) simulations opened the way to a most relevant field
of investigation but also revealed, at that time, an irremediable
discrepancy between predicted and observed periods. The issue was
further investigated by several authors (see, e.g., Caputo et al.
1987; Lee et al. 1990; Caputo et al. 1993; Lee et al. 1994; 
Demarque et al. 2001) who discussed several important
features such as the dependence of the RR Lyrae absolute magnitude
$M_V$(RR) on metallicity and HB morphology, the period-shift at
fixed effective temperature, and the {\it second} parameter that,
in addition to metal abundance, influences HB morphology.

More recently, both the stellar evolution and the pulsation theory
have greatly improved, whereas no exhaustive investigations have
appeared in the relevant literature discussing synthetic period
distributions of RR Lyrae stars at various metal content. On this
basis, we planned to revisit this problem on modern grounds,
presenting in this first paper a discussion of the observational
scenario concerning RR Lyrae periods in Galactic globulars. This
is intended as a first step in a more extended investigation we
are performing on the basis of the predictions from updated
pulsation models collated with SHB simulations. The theoretical
framework will be presented and discussed in a series of
forthcoming papers, together with the analysis of RR Lyrae stars
in individual clusters.

In the following, we will take advantage of the careful catalogue
of RR Lyrae data made available by C. Clement (see Clement et al.
2001) to present an up-to-date compilation of period frequency
histograms for RR Lyrae stars in a selected sample of Galactic
globular clusters. The last collection of these
histograms date back to more than fifteen years ago (Castellani \&
Quarta 1987: hereinafter CQ87) whereas in the meantime
observational data have been significantly improved. In the next
section we will present these histograms, discussing some relevant
evidence to the light of current theoretical predictions. The
behavior of mean periods will be investigated in Sect. 3, with
particular attention to the sample of clusters with a
statistically significant RR Lyrae population. Sect. 4 will deal
with the origin of the Oosterhoff dichotomy, whereas brief remarks
will close the paper.

\section{The Period frequency histograms}

To enlighten the role of period frequency histograms let us remind
that both on theoretical and observational ground we know that
within the instability strip one finds either first overtone (RRc)
and fundamental (RRab) pulsators, with only first overtone or
fundamental mode being unstable at the larger or at the lower
temperatures, respectively. Moreover, the period of a RR Lyrae
pulsator can be quite firmly linked to the pulsator evolutionary
parameters (mass, luminosity and effective temperature), with
periods becoming larger when the luminosity increases or the
effective temperature and/or the stellar mass decrease. Thus, in
each cluster the observed distribution of periods is  a signature
of the distribution of the mass, luminosity and
effective temperature of the
pulsators.

As early stated by Stobie (1971), the correct way to investigate
such a distribution is to transform the periods so that all the
variable are pulsating in one mode, say the fundamental one.
According to the pulsation models by van Albada \& Baker (1973),
this can be  done with good accuracy by simply adopting for c-type
pulsators the "fundamentalized" period log$Pf$=log$Pc$+0.13, as
already given in CQ87. More recent computations (Bono et al. 1997,
but see also Marconi et al. 2003 [Paper II]) support such a
procedure, with an uncertainty on the fundamentalized periods not
larger than $\delta$log$P_f= \pm$0.005. The advantage of using
fundamentalized periods is that in this way both RRc and RRab
periods follow just the same pulsation relation, so that the
observed distribution of fundamentalized periods is directly
reflecting the actual distribution of star masses, luminosities
and effective temperatures across the whole instability strip.

For our compilation we selected from Clement's catalogue the
sample of Galactic globular clusters with more than 12 pulsators
with well determined period and pulsation mode, assumed as
representative of clusters with a not occasional occurrence of RR
Lyrae stars.  Presently one finds 32 clusters in this class,
against 28 in the previous CQ87 paper, with a total of 1545
variables and an overall increase in the number of pulsators
larger than 30\%. Figure 1 shows the period frequency histograms
for the quoted cluster sample, as arranged in order of increasing
metallicity [Fe/H] as given in the 1999 update of Harris (1996)
compilation. In the figure, the black areas give the contribution
of fundamentalized c-type periods. As well known, one finds that
first overtone pulsators contribute to the shortest portion of the
period histograms, according to their location on the hotter side
of the instability strip.

In  Fig. 1 one may notice  that the range of periods covered by M3
pulsators can be taken as roughly representative of the behavior
of pulsators in the large majority of clusters, independently of
the metallicity, with only a minor suggestion for a possible shift
toward smaller periods at the larger metallicities. Here we do not
give too much relevance to such an evidence, which will be better
discussed in forthcoming papers relying on SHB procedures. We only
notice that a constant minimum period over the metallicity range
$Z$=0.0001 to 0.001 is the one predicted on the basis of HB
stellar models, since an increase in metallicity is expected to
decrease both luminosity and mass in such a way to keep periods
roughly constant, or at the most to decrease periods by
$\Delta$log$P\sim$ 0.01 (see Bono et al. 1995), whereas a slight
decrease of period is expected when further increasing the metal
content above $Z$=0.001. The above agreement between these
predictions and observation supports the theoretical dependence of
evolutionary parameters on metallicity.

By looking in Fig. 1 one finds two histograms which clearly fall
outside the M3 period range, as given by  M15 (and perhaps
NGC5466), with the occurrence of a small number of very short
period (log$Pf\le-$0.5) pulsators, and  by NGC6441, where periods
appears clearly shifted toward larger values. As for M15, it
appears hard to account for the  shortest periods moving either
the star evolutionary parameters or the instability boundaries
within reasonable values. However, these pulsators have been found
(Butler et al. 1998) in the crowded central cluster region, with
only a partial coverage of the light curve. If these periods will
be confirmed, this could be regarded as an evidence for second
overtone instability, an occurrence not foreseen by current
theoretical investigations. As for NGC6441, a very metal rich
cluster with the  HB reaching quite hot effective temperatures
(Rich et al. 1997), the peculiar large periods of RR Lyrae stars
cannot be understood within the current evolutionary scenario (see
Sweigart \& Catelan 1998), requiring "ad hoc" assumptions as we
will discuss in a forthcoming paper on the basis of synthetic HB
simulations. We note that anomalously large RR Lyrae
periods have been found also in NGC6388, another metal rich
cluster with peculiarly extended HB (Pritzl et al. 2002). Given
the small numbers of well studied variables (4 RRab and 7 RRc),
this cluster is not included in the present study.

\section{Mean periods}

As often adopted in the current literature, the behavior of RR
Lyrae in globular clusters can be discussed in terms of mean
periods for fundamental and/or first overtone pulsators. However,
when approaching such an issue, one has to bear in mind that
observational data are affected by statistical fluctuations. In
CQ87 we have already approached the problem, showing that typical
statistical uncertainties on mean (fundamental or fundamentalized)
periods in cluster with 20, 40 or 80 RR Lyrae stars are of the
order of $\delta$log$P\sim \pm$ 0.03, 0.02 or 0.01, respectively.
Moreover, one has to care about the occurrence of clusters with
only a partial population of the strip, where mean periods are
governed by this peculiar occurrence. According to these warnings,
to discuss mean periods we selected from the sample of 32 clusters
those having no less than 10 RRab and 5 RRc with well determined
period, fixing however the attention on clusters with more than 40
RR Lyrae stars from which one expects more stringent indications.

Table 1 ranks these clusters in order of increasing Harris's
metallicity, giving for each cluster, left to right, the NGC/IC and
(if existing) the Messier number, the iron abundance according to
Harris or in the Carretta \& Gratton (1997) scale, the HB type
(Lee 1990, from Harris catalogue), and the total number of RR Lyrae
stars found in the cluster, followed by the numbers of pulsators
recognized as RRab or RRc, including in the latter class the
suspected or suggested second overtone pulsators. The following
two columns give the mean (logarithmic) period for the fundamental
pulsators ($<$log$Pab>$) and for the whole sample ($<$log$Pf>$),
after fundamentalising the RRc periods. Last column finally gives
the cluster Oosterhoff type.

\begin{table*}
\begin{center}
\caption[]{Selected quantities for Galactic globular clusters with
no less than 10 RRab and 5 RRc. Bold characters with asterisks
mark the restricted sample of clusters with more than 40 RR Lyrae stars.
\label{tab1}}

\begin{tabular}{c c c c c c c c c c c c c}

\hline \hline

NGC & Messier & [Fe/H]$_H$  & [Fe/H]$_{CG}$  & HB type & RR  & RRab   & RRc &
$<$log$Pab>$  & $<$log$Pf>$ & Oo-type\\

\hline

NGC6341 &M92 &-2.29& -2.16 & 0.91 & 17 & 11 & 6 &-0.201&-0.244& II\\
\bf *NGC7078&M15&-2.25&-2.12& 0.67 & 105& 39 & 57&-0.192&-0.258& II\\
NGC5466 &    &-2.22& -2.1  & 0.58 & 22 & 13 & 7 &-0.193&-0.229& II\\
NGC2419 &    &-2.12& -1.97  &0.86  &34  &25 & 7 &-0.185&-0.207& II\\
\bf *NGC4590 &M68& -2.06 &-1.99  &0.17  &44  &14 &28 &-0.211&-0.264& II\\
\bf *NGC5024 &M53& -1.99& -1.81  &0.81  &59  &29 &29 &-0.190&-0.260& II\\
NGC6656 &M22 &-1.64 &-1.48&  0.91 & 26 &10 & 9  &-0.201&-0.257& II\\
\bf *NGC7006&    & -1.63 &-1.35& -0.28  &63 &50  &8  &-0.245&-0.259& I\\
NGC7089 &M2&  -1.62 &-1.48&  0.96  &30 &18 &12  &-0.168&-0.241& II\\
\bf *IC4499 &   &  -1.60 &-1.46&  0.11  &98 &63 &34  &-0.239&-0.270 & I\\
\bf *NGC6715 &M54& -1.59& -1.25 & 0.87&  67& 50&  5  &-0.240&-0.250 & I\\
\bf *NGC3201&   &  -1.58 &-1.23&  0.08&  81 &71&   5 &-0.256&-0.264&I\\
\bf *NGC5272 &M3&  -1.57& -1.34&  0.08& 235&173&  51& -0.254&-0.274&I\\
\bf *NGC6934 &   & -1.54 & -1.30      &0.25 & 79& 68&  10 &-0.245&-0.264&I\\
NGC6584& &    -1.49 &-1.30& -0.15 & 42 &33&   7 &-0.254&-0.276&I\\
NGC6229&  &   -1.43 &-1.30  &0.24 & 38 &30   &8 &-0.259&-0.282 &I\\
\bf *NGC6402 &M14& -1.39 &-1.27&  0.65 & 55& 40&  14& -0.246&-0.276&I\\
NGC1261&   &  -1.35 &-1.08& -0.71&  19& 13&   5 &-0.257&-0.284&I\\
\bf *NGC5904 &M5&  -1.29& -1.11&  0.31& 132  &88& 37& -0.261&-0.294&I\\
\bf *NGC6266 &M62& -1.29& -1.04&  0.32&  74 & 62&  12&-0.268&-0.289&I\\
NGC1851&  & -1.22 &  -1.05& -0.36&  29 & 21&  8& -0.246&-0.284&I\\
NGC6121 &M4& -1.20&  -1.19& -0.06&  39&  30&  9 &-0.272&-0.304&I\\
NGC6723&  & -1.12&   -0.96& -0.08&  29&  23&  5& -0.270&-0.294&I\\
NGC6171& M107& -1.04&-0.91& -0.73&  22&  15&  7& -0.272&-0.317&I\\
NGC6362 &  &-0.95&   -0.96& -0.58&  36&  18 &18& -0.265&-0.333&I\\
\bf *NGC6441 &  &-0.53&         & 0.00&  43 & 27& 16& -0.128&-0.191&?\\
\hline \hline
\end{tabular}
\end{center}
\end{table*}

The position of this sample of "RR Lyrae rich clusters" within the
family of Galactic globular clusters is shown by filled symbols in
Fig. 2, where we report the Lee's diagram (HB type against
metallicity) for all the globulars with the HB type listed
in the Harris
catalogue, by adopting Harris metal abundances.

 Data in this figure,
though affected by observational uncertainties, deserve some
comments. Taken at their face value, the data reveal the well
known occurrence of a "second" parameter beside [Fe/H] governing
the HB morphology, since clusters with metallicity around
[Fe/H]$\sim -$1.4 may have HB type ranging from +1 to $-$1, i.e.,
from very blue to completely red HB population. Moreover, as
already discussed in CQ87, one finds that the "RR Lyrae rich"
clusters (filled symbols) form two separate groups, with no
cluster in the metallicity range [Fe/H] $\sim -$1.6 to $-$2.0.

\begin{figure}
\resizebox{\hsize}{!}{\includegraphics{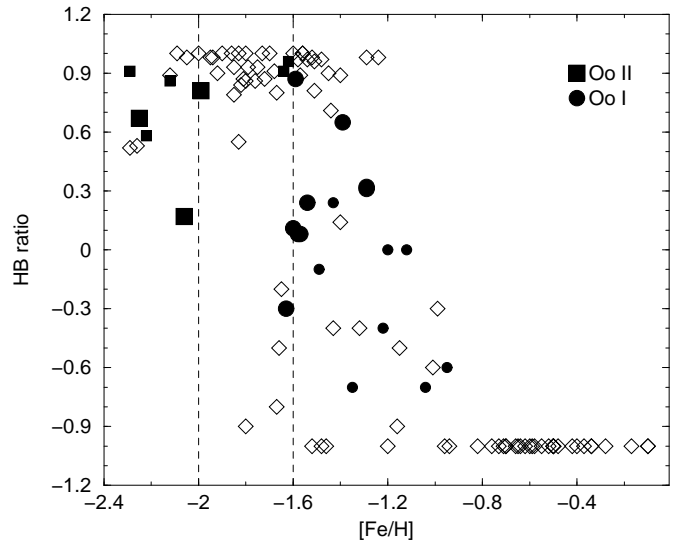}}
\caption{HB type versus cluster [Fe/H] form Harris catalogue. Filled symbols
show the sample of RR rich clusters, the largest symbols
indicating clusters with more than 40 RR Lyrae variables.}
\end{figure}

Inspection of data in Table 1 shows that such a dichotomy is
preserved even adopting the alternative Carretta \& Gratton (1997)
metallicity scale. Below [Fe/H] $\sim -$2.0 one finds only OoII
type clusters, whereas above [Fe/H] $\sim -$1.6 one finds the
whole family of OoI clusters plus two OoII clusters (M2 and M22)
placed just at the lower metallicity edge. The occurrence of these
two "anomalous" OoII clusters is among the observational evidences
to be explained by a theoretical scenario for RR pulsators in
globulars. Here we only notice that both clusters are not very
rich in variables, presenting a blue HB population.

Concerning the mean periods, the upper panel in Fig. 3 shows the
mean fundamental period $<$log$Pab>$ of OoI (filled dots) and OoII
(filled squares) clusters in Table 1, as a function of the cluster
metallicity, while the lower panel gives the distribution for the
mean fundamentalized period $<$log$Pf>$. One finds that the
stringent selection of clusters with more than 40 RR Lyrae stars
allows to derive from these well known and well discussed diagrams
(see, e.g., Clement et al. 2001) some relevant points. From data in
the upper panel, one finds that $<$log$Pab>$ follows reasonably
strictly the cluster metallicity, with a clear separation between
OoII and OoI types which arrange around $<$log$Pab>$ $\sim -$0.20
and $-$0.25, respectively, and with an indication for further
decreasing periods at the metal rich tail. One has to notice the
sudden dichotomy separating OoII from OoI clusters even for
clusters with similar metallicity such the couples M22-M2 (OoII)
and NGC7006-IC4499 (OoI). The peculiar cluster NGC6441, as marked
in Fig. 3 with a triangle, will be not further discussed in this
paper.

However, the most interesting feature appears in the lower panel
of the figure, where one finds again a correlation between
$<$log$Pf>$ and the cluster metallicity, but with the additional
evidence that the mean fundamentalized period is constant for
metal content lower that [Fe/H]=$-$1.6, so that a common value is
"connecting" OoII and OoI clusters across the metallicity gap, and
with again the indication for a period decrease at the metal rich
tail. As a whole, data in Fig. 3 suggest that $<$log$Pf>$ in RR
Lyrae rich clusters does closely follow the cluster metal content,
with the mean periods in RR Lyrae less populated clusters being
spread off because of statistical fluctuation and/or strong
inhomogeneity in the strip population.

\begin{figure}
\resizebox{\hsize}{!}{\includegraphics{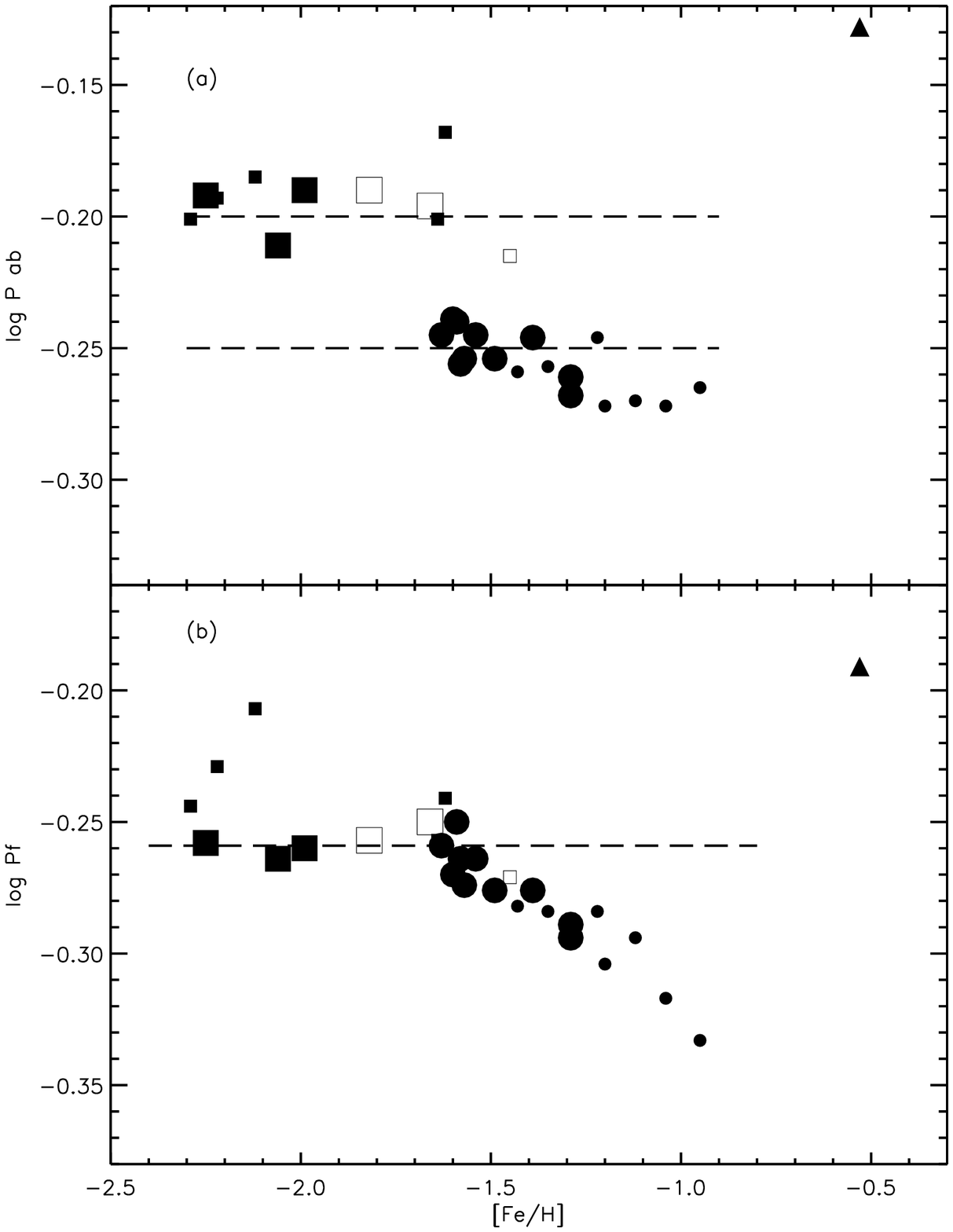}}
\vspace*{0.4truecm} \caption{Mean periods of fundamental (a) or
fundamentalized  (b) RR Lyrae pulsators as a function of the
cluster metallicity for the sample of 26 "RR rich" globular
clusters in Table 1, with filled symbols as in Fig. 2. Open
symbols refer to the three selected groups of "RR poor" clusters
in Table 2 (see text). The triangle depicts the "anomalous"
cluster NGC6441.}
\end{figure}

\begin{table*}
\begin{center}
\caption[]{RR Lyrae poor clusters  as grouped according to their
metallicity. Bold characters with asterisks gives the cumulative
properties of the three groups. \label{tab2}}

\begin{tabular}{c  c c c c c c c c c c c}

\hline \hline

NGC & Messier & [Fe/H]$_H$   & HB type &  RRab &
RRc &
$<$log$Pab>$  & $<$log$Pf>$ & Oo-type\\

\hline
NGC4833 &    &-1.79&   0.93 & 7 & 8 & & & \\
NGC5897 &    &-1.80&   0.86 & 3 & 8 & & & \\
NGC6809 & M55&-1.81&   0.87 & 4 & 9 & & & \\
NGC4147 &    &-1.83&   0.55 & 6 & 10& & & \\
NGC2298 &    &-1.85&   0.93 & 1 & 3 & & & \\
NGC5824 &    &-1.85&   0.79 & 8 & 1 & & & \\

\vspace*{0.4truecm}
\bf *Group 1&  &-1.82 $\pm$ 0.03&  $\ge$ 0.55& 29& 39&-0.190&-0.257&II\\

NGC1904 &M79 &-1.57&   0.89 & 2 & 1 & & & \\
NGC5986 &    &-1.58&   0.97 & 7 & 1 & & & \\
NGC5286 &    &-1.67&   0.80 & 9 & 5 & & & \\
NGC6333 & M9 &-1.72&   0.87 & 8 & 9 & & & \\
NGC6093 &M80 &-1.75&   0.93 & 4 & 2 & & & \\

\vspace*{0.4truecm}
\bf *Group 2&  &-1.66 $\pm$ 0.09&  $\ge$ 0.80& 30& 18&-0.196&-0.250&II\\

NGC288 &     &-1.24&   0.98 & 1 & 1 & & & \\
NGC6235 &    &-1.40&   0.89 & 2 & 1 & & & \\
NGC6626 &M28 &-1.45&   0.90 & 8 & 2 & & & \\
NGC6681 &M70 &-1.51&   0.96 & 1 & 1 & & & \\
NGC7492 &    &-1.51&   0.81 & 1 & 2 & & & \\
NGC6205 &M13 &-1.54&   0.97 & 1 & 4 & & & \\

\vspace*{0.4truecm}
\bf *Group 3&  &-1.45 $\pm$ 0.07&  $\ge$ 0.81& 14& 11&-0.215&-0.271&II\\
\hline \hline
\end{tabular}
\end{center}
\end{table*}

However, following the referee's suggestion, one can explore the
behavior of RR Lyrae poor clusters in the "metallicity gap"
$-$2.0$<$[Fe/H]$<-$1.6 between OoI and OoII types, by looking at
the cumulative contribution of the known variables in clusters of
comparable metal content. Our original intention was to group the
clusters according to both their metallicity and HB type. However,
one finds that in the quoted metallicity range all the clusters
containing well studied RR Lyrae stars, but Ruprecht 106, have
blue HB populations, with HB types larger than +0.5. As for
Ruprecht 106 ([Fe/H]=$-$1.67, HB type $-$0.82), it has repeatedly
found to be younger than other galactic globulars (see, e.g.,
Buonanno et al. 1993) and appears as a peculiar OoI type cluster,
with 13 RRab ($<$log$Pab>$=$-$0.24) and no RRc variables.
Excluding this object, the remaining clusters have been collected
in two groups according to their metallicity,  as shown in Table
2.

It turns out that both groups behave as typical OoII type
clusters, with mean $<Pab>$ and $<Pf>$ periods in good agreement
with the values found in RR Lyrae rich OoII type clusters. Adding
these results (open squares) to the data already plotted in the
previous Fig. 3, one finds an increasing evidence for a sudden
dichotomy of  $<$log$Pab>$ around [Fe/H]$\sim$ -1.6, together with
a beautiful support to the constancy  of $<$log$Pf>$ passing from
OoII to OoI type clusters.

We have extended the analysis also to the remaining RR Lyrae poor
clusters with blue HB morphology (HB type $\ge$0.80) and even larger metal abundance.
As listed in Table 2, this third group characterized by OoI-like
metal abundance ($<$[Fe/H]$>\sim-$1.45) discloses OoII features as
the mean $<Pab>$ period is concerned, providing additional proofs
that the Oosterhoff dichotomy is not a matter of metal abundance
only, but is also governed by the HB morphology.

\section{The Oosterhoff controversy}

As recalled in the introduction to this paper, since  long time we
are facing the puzzling evidence for the dichotomized behavior of
the mean period of fundamental variables in Galactic globular
clusters. As for the origin of such a dichotomy, the larger
$<Pab>$ observed in  OoII type clusters could be the direct result
of common larger pulsator periods, as suggested by Sandage (1982).
However, according to an early hypothesis by van Albada \& Baker
(1973), OoII type clusters could have longer mean period $<Pab>$
for the simple reason that the low period fundamental pulsators
observed in OoI clusters are now pulsating in the first overtone,
and thus are not contributing to the average of RRab periods.
Further pulsational calculations (Stellingwerf 1975) supported
such a picture, disclosing the occurrence within the instability
strip of an "OR" zone, where the stars can pulsate either in the
fundamental or in first overtone mode, separating the regions
where only the first overtone (at larger effective temperatures:
FO region) or only the fundamental mode (at lower effective
temperatures: F region) are stable (see also Paper II and
references therein.

Based on the evidence presented in the previous sections,  we are
now in the position to draw some firm conclusions. As a first
point, one may notice that the already quoted good constancy of
the fundamentalized period range in both OoI and OoII clusters
indicates that the width in temperature of the strip remains
honestly constant, ruling out in the meantime the possibility that
OoII pulsators have simply larger luminosity and, thus, larger
periods than OoI pulsators. In that case, given the difference
$\Delta <$log$Pab>$=0.06 between the two prototypes M15 (OoII) and
M3 (OoI), one would indeed expect the whole period histogram of
M15 shifted by the same amount with respect to M3, and this is not
the case. Such a conclusion is reinforced by data plotted in Fig.
3, showing that  OoI and OoII clusters at [Fe/H] $\leq -$1.6 have
quite similar $<$log$Pf>$. This is exactly what expected if the
Oosterhoff dichotomy is produced by a change in the pulsation
mode, which obviously does not affect the fundamentalized periods.

One can find several additional evidences supporting these
conclusions. The direct comparison of the period frequency
histograms in M3 and M15, as given in Fig. 4, discloses the
occurrence of an intermediate range of fundamentalized periods
where a substantial amount of stars is pulsating as first overtone
in M15, but as fundamental pulsators in M3.  We conclude that the
variation in the pulsation mode, even if not the unique cause,
does play a dominant role in governing the Oosterhoff dichotomy.

\begin{figure}
\resizebox{\hsize}{!}{\includegraphics{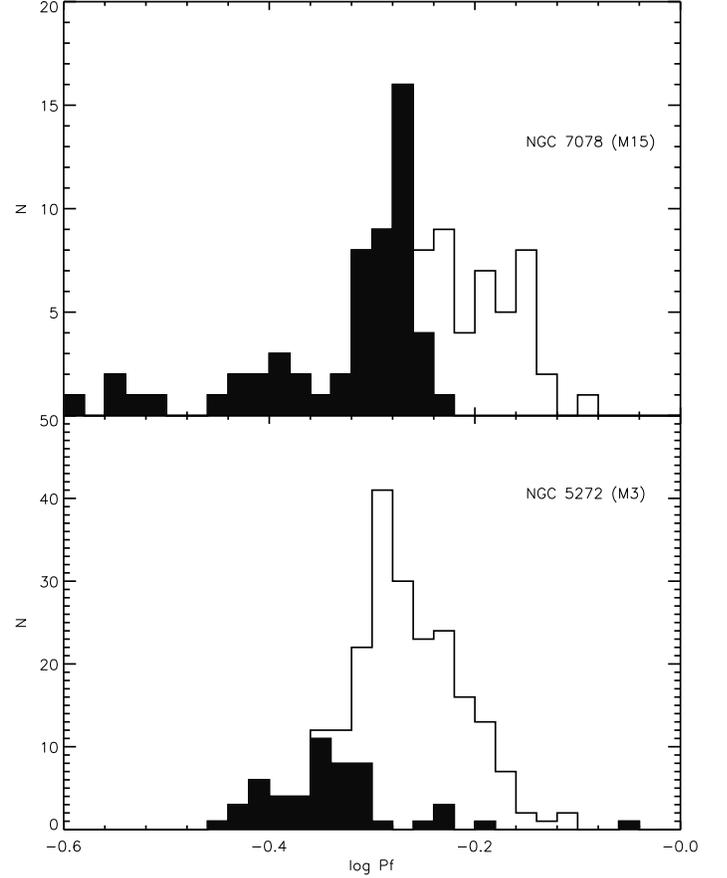}}
\vspace*{0.4truecm} \caption{The period frequency histograms for
M3 and M15. Black bars show the contribution by fundamentalized
first overtone pulsators.}
\end{figure}

Moreover, an overall picture of the Oosterhoff dichotomy must
include the well-known variation in the relative abundance of
c-type RR Lyrae stars, as given by the number ratio Nc/Ntot, where
Ntot=Nab+Nc. As shown in Fig. 5, the most RR Lyrae-rich clusters
show indeed that the $<Pab>$ dichotomy (dashed lines) is
accompanied by a bimodality (arrows) in the ratio Nc/Ntot which
cannot be ascribed to a simple period variation. On the other side,
moving the transition from RRab to RRc variables toward lower
temperatures (i.e., transforming fundamental in first overtone
pulsators) has the twofold effect of increasing $<Pab>$ and
increasing the relative number of first overtones. If $<logPf>$ is
plotted against Nc/Ntot, the dichotomy in periods obviously
disappears, while remains the signature of the variation in the
transition temperature, as given by the larger amount of first
overtone pulsators in OoII type clusters.

\begin{figure}
\resizebox{\hsize}{!}{\includegraphics{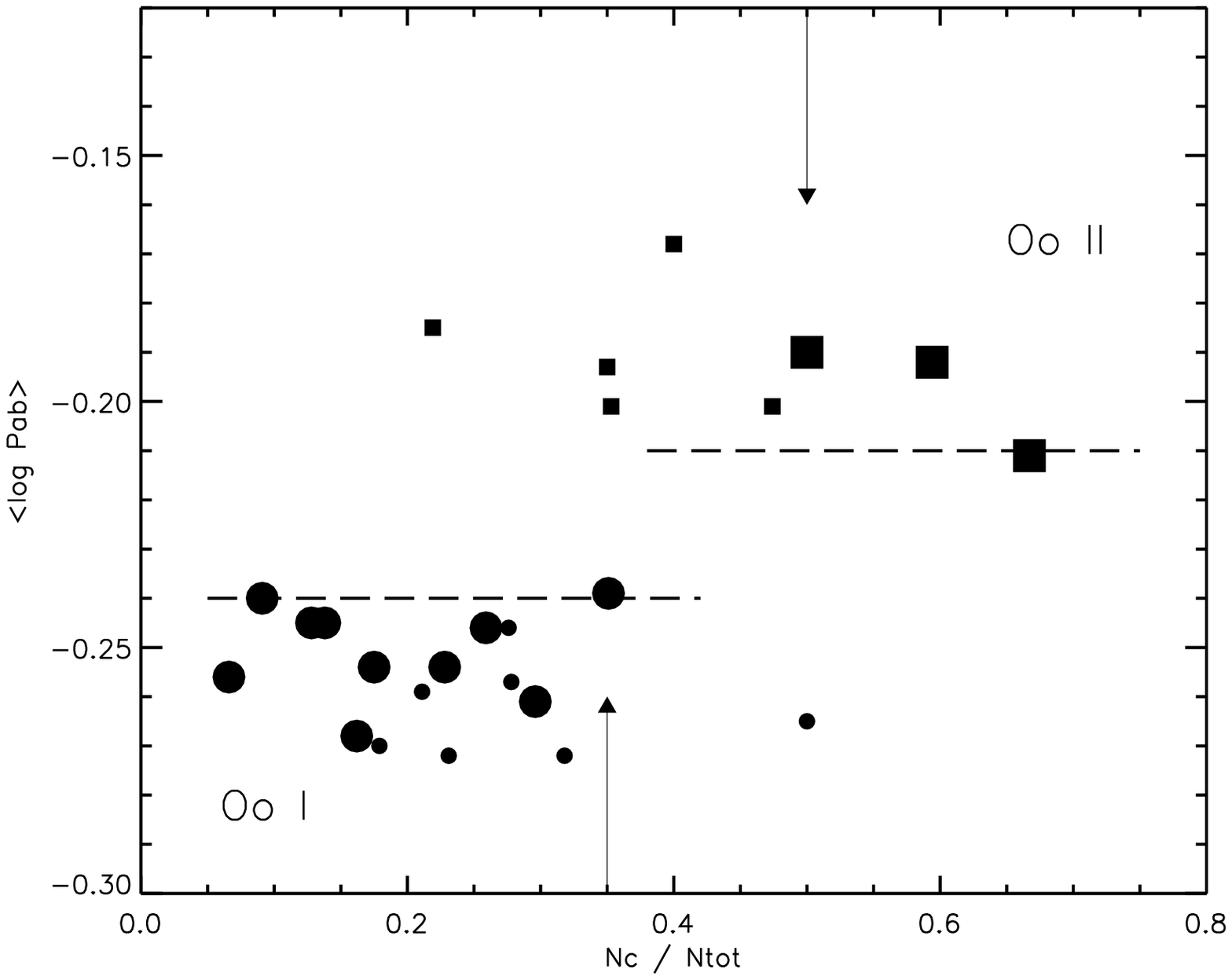}}
\vspace*{0.4truecm}
\caption{The bimodal distribution of mean periods of RRab
variables as a function of
the number ratio of first overtone pulsator (Nc)
to the total number of RR Lyrae (Ntot).
Symbols as in previous figures.}
\end{figure}

One may finally reinforce this conclusion on the basis of simple
theoretical considerations. In a uniformly populated instability
strip, if - as expected - pulsator luminosities and masses can be
assumed as honestly constant, the difference
$\Delta$log$P$=$<$log$Pab>- <$log$Pf>$ depends only on the
difference log$Te$(FOBE)$-$log$Te$(TR) between  the temperature of
the First Overtone Blue Edge (FOBE: instability blue boundary) and
the transition temperature, independently of the pulsator
luminosity.  On this basis, the significative variation of
$\Delta$log$P$ between OoI and OoII clusters appearing in the
previous Fig. 4 has to be ascribed to a variation in the
difference  log$Te$(FOBE)$-$log$Te$(TR).

On a quantitative point of view, at constant mass and luminosity,
one has log$P$=const$-$3.36log$Te$ (Paper II, but see also van
Albada \& Baker 1973). It follows

\vspace{0.2cm}

\noindent $<$log$Pab>-<$log$Pf>$=1.68[log$Te$(FOBE)$-$log$Te$(TR)]

\vspace{0.2cm}

\noindent Adopting from the most RR Lyrae rich clusters in Fig. 3
\vspace{0.2cm}

\noindent $<$log$Pab>-<$log$Pf>\sim$ 0.02 (OoI) or $\sim$0.06
(OoII)

\vspace{0.2cm}
\noindent one derives
\vspace{0.2cm}

\noindent log$Te$(FOBE)$-$log$Te$(TR)$\sim$ 0.01 (OoI) or 0.04
(OoII) \vspace{0.2cm}

\noindent in reasonable  agreement with theoretical predictions
concerning the range of temperatures covered by the FO region or
the FO plus the OR region, respectively (see Fig.1 in Bono et al.
1997).

\begin{figure}
\resizebox{\hsize}{!}{\includegraphics{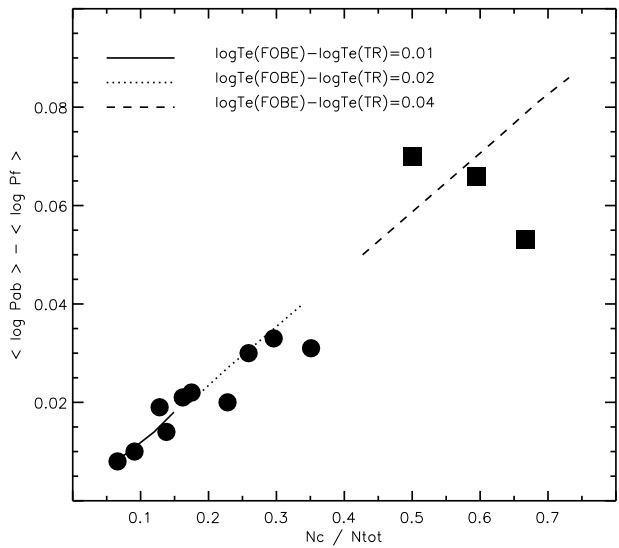}}
\vspace*{0.4truecm} \caption{ The observed $<$log$Pab>-<$log$Pf>$
for the most RR Lyrae rich globular clusters (symbols as in Fig. 2),
as a function of the number ratio of first overtone pulsator (Nc)
to the total number of RR Lyrae (Ntot). The three segments show
the theoretical predictions for the labelled assumption on the
difference  log$Te$(FOBE)$-$log$Te$(TR) when varying HB morphology
(see text). }
\end{figure}

This scenario is supported by numerical experiments based on
synthetic HB procedures, allowing to take into account realistic
variations in mass and luminosity, as well as the effects of not
uniform strip populations. To this purpose we made use of
HB models as computed in the
case of inefficient element sedimentation and adopting  the mixing
length  parameter $l$=2.0$H_p$, as calibrated on globular cluster
Red Giant Branches (Cassisi et al. 1999). Figure 6 shows the predicted
$\Delta$log$P$ and Nc/Ntot ratios inferred by synthetic
simulations with  HB type in the range of $\sim -$0.75 to
$\sim$+0.75, as evaluated under selected assumptions about the
difference between $Te$(FOBE) and the transition temperature
($\Delta$log$Te$=0.01, 0.02, and 0.04). One finds that for each
given $\Delta$log$Te$,  different HB star distributions can move
the predicted $\Delta$log$P$ and Nc/Ntot ratio only by a limited
amount. Conversely, observational data run against a common
$\Delta$log$Te$, with first overtone pulsators in OoI clusters
covering an effective temperature interval  of the order of
log$Te$(FOBE)-log$Te$(TR)$\sim$ 0.01-0.02, whereas in OoII
clusters they suggest a much larger value, of the order of $\sim$
0.04.

\section{Final remarks}

In this paper we have revisited observational data concerning RR
Lyrae pulsators in Galactic globular clusters, presenting period
frequency histograms for variables in  the 32 clusters having more
than 12 RR Lyrae with well recognized period and pulsation mode.
Inspection of the histograms shows that the range of
fundamentalized periods covered by variables in a cluster remain
roughly constant in varying the cluster metallicity all over
the cluster sample, with only two exceptions and with a minor
suggestion for a possible shift toward smaller periods at the
larger metallicities. We conclude that the width in temperature of
the instability strip appears largely independent of the cluster
metallicity, whereas  the evidence that  the period 
distribution is not affected by the predicted variation of  pulsators
luminosity with metal abundance can be understood in terms of a
correlated variation in the pulsator mass, as predicted by current
evolutionary theories.

We discuss mean periods, selecting from the sample of 32 clusters
26 clusters having no less than 10 RRab and 5 RRc, but paying
particular attention to clusters with more than 40 RR Lyrae stars,
as the only ones with a sufficient statistical significance.  One
finds the well known Oosterhoff dichotomy in
the mean $<Pab>$ periods, together with similarly clear evidence for
a continuous behavior
of the fundamentalized mean period $<Pf>$ in passing from
OoII to
OoI clusters, as well as for its constancy between the two groups,
at fixed metal abundance.

On this basis the origin of the Oosterhoff
dichotomy is briefly discussed, presenting  evidences against a
strong dependence of the HB luminosity on the metal content, as
originally suggested by Sandage (1982) and more recently by
D'Antona et al. (1997). On the contrary, the continuity of
fundamentalized period, the comparison of the period frequency
histograms in M3 and M15, the bimodal distribution of the relative
abundance of first overtone pulsators and, last but not least, the
observed differences between mean  fundamental and fundamentalized
periods,  all agree in indicating the dominant occurrence of a
variation in the pulsation mode, with variables in the OR zone
pulsating in the fundamental or in the first overtone
 mode in OoI and
OoII type clusters, respectively.

\begin{acknowledgements}

We are indebted to our anonymous referee for pertinent comments
and suggestions that improved the first version of the paper. We
gratefully acknowledge G. Bono, R.F Butler, H. Smith and A.R.
Walker for helpful suggestions. Financial support for this work
was provided by the Ministero dell'Istruzione, dell'Universit\`a e
della Ricerca (MIUR) under the scientific project ``Stellar
observables of cosmological relevance'' (V. Castellani \& A.
Tornamb\`e, coordinators).

\end{acknowledgements}


%


%









\end{document}